\documentclass[prd, aps, twocolumn, showkeys, showpacs, 10pt]{revtex4-1}

\usepackage{amssymb}

\def\bea{\begin{eqnarray}}
\def\eea{\end{eqnarray}}
\def\be{\begin{equation}}
\def\ee{\end{equation}}
\def\ba{\begin{array}}
\def\ea{\end{array}}
\def\nn{\nonumber}

\begin{document}

\title{Bigravitational inflation}
\author{Vicente Atal,$^{a}$ Luis E. Campusano,$^{a}$ and Gonzalo A. Palma$^{b}$}

\affiliation{
$^{a}$Departamento de Astronom\'ia, FCFM, Universidad de Chile, Casilla 36-D, Santiago, Chile\\
$^{b}$Departamento de F\'isica, FCFM, Universidad de Chile, Casilla 487-3, Santiago, Chile
}

\begin{abstract}
We study the realization of cosmic inflation in bigravity theories. By analyzing the evolution of scalar, vector, and tensor perturbations in de Sitter-like spacetimes, we find strong stability constraints on the class of viable vacua offered by these theories. More specifically, the only stable de Sitter vacua contain two nondecoupled gravitons (one of which is massive) with different maximal propagation speeds. We derive an effective theory for the massless graviton, which is found to propagate at an intermediate speed, limited by the two maximal values.  For inflation, while the spectrum of density perturbations remains nearly scale invariant, the power spectrum of tensor modes is found to  depart  from the usual prediction found in standard slow-roll inflation.  In particular, both the tensor to scalar ratio $r$ and the spectral index of tensor modes $n_T$ receive sizable contributions from the couplings of the theory, leading to specific signals that may be tested in future cosmological probes of CMB polarization.
\end{abstract}

\date{\today}
\keywords{Cosmic inflation,  Modified theories of gravity, Bigravity theories}
\pacs{98.80.Cq, 04.50.Kd, 04.30.-w}

\maketitle


\section{Introduction}

Our present view of the Universe relies entirely on the validity of Einstein's general relativity (GR), which continues to be our best explanation to all known gravitational phenomena. 
Our great confidence in GR is reflected in our acceptance of its inference for the existence of dark matter and dark energy, based on several astronomical and cosmological observations, including galaxy rotation curves, supernovae redshift-distance relation, CMB, and large scale structure.
Despite its undisputed success, the lack of a deeper insight on the nature of these {\it dark substances} still raises the question as to whether GR constitutes the correct theoretical framework by which the gravitational interaction should be addressed. 

Instead, GR might turn out to be an effective description of gravity valid at intermediate scales that needs to be completed at both UV and IR scales, leading to a more fundamental theory~\cite{Clifton:2011jh}.  One concrete possibility is put forward by bigravity \cite{Isham-1}, which asserts the existence of a second spin-2 particle ---in addition to the usual graviton--- nontrivially modifying the long-range action of gravity. Recent developments have made clear that bigravity models imply significant but consistent departures from GR at long and short wavelengths~\cite{Damour:2002ws, Blas:2005yk, Clifton:2010hz}. As a consequence, these models offer an alternative view of phenomena such as dark matter and dark energy, leading to interesting prospects for future tests on gravity~\cite{Banados:2008fi, Banados:2008fj}. 

A key test on bigravity is whether it is able to provide an explanation for the origin of primordial density fluctuations~\cite{Lyth:1998xn} as observed in cosmic microwave background (CMB)~\cite{Komatsu:2010fb} and large scale structure measurements~\cite{Sanchez:2005pi}.  The purpose of this note is to address this question by analyzing the realization of cosmic inflation in bigravity models. We start by first studying the evolution of perturbations on the class of de Sitter backgrounds offered by these theories. We show that only a restricted family of de Sitter vacua is stable under perturbations, characterized by the fact that its two gravitons remain coupled at wavelengths comparable to the horizon. Then, by assuming that the de Sitter geometry evolves adiabatically towards a Minkowski vacuum (a quasi-de Sitter state), we find that the power spectrum of tensor modes receives contributions that makes it differ from the usual prediction encountered in slow-roll models of inflation, offering a unique opportunity to test bigravity, involving measurements of CMB polarization.

\section{The model}

We begin our discussion by introducing the basic setup to be studied, namely, a system consisting of two metric fields $g_{\mu \nu}$ and $q_{\mu \nu}$ with inverse fields $g^{\mu \nu}$ and $q^{\mu \nu}$, respectively.  The action describing this system is given by
\bea
S_{\rm BG} &=& \frac{1}{L^2}\!\! \int \!\! d^4 x \! \left[  \sqrt{-g} \! \left(\frac{R_g}{2} -  \frac{\lambda_g}{L^2} \right) +  \sqrt{-q} \! \left( \frac{R_q}{2} -  \frac{\lambda_q}{L^2} \right)  \right]  \nn \\
&& -  \frac{\beta}{2 L^4}  \int \!\! d^4 x (-q)^{u}(-g)^{v}  g_{\mu \nu} q^{\mu \nu}   ,   \label{bi-gravity-action-1}
\eea
where $R_g$ and $R_q$ are the Ricci scalars constructed from $g_{\mu \nu}$ and $q_{\mu \nu}$, and $\lambda_g$ and $\lambda_q$ play the role of cosmological constants for each sector. The length scale $L$ is introduced to make all parameters dimensionless, and may be taken as the fundamental length scale of gravity. The constant $\beta$ couples $g$ and $q$ with the help of a mixed volume element $d^4 x (-q)^{u}(-g)^{v} $ with $u + v = 1/2$.  The presence of $\beta$ breaks the diff$^2$ symmetry of the noninteracting theory down to diff \cite{Boulanger:2000rq}, making (\ref{bi-gravity-action-1}) to be invariant only under simultaneous gauge transformations of both metrics. We add to action (\ref{bi-gravity-action-1}) an interaction term first proposed in \cite{Isham-1}
\be
S_{\rm int} =  \frac{\kappa}{2 L^4} \!\! \int \!\! d^4 x (-q)^{u}(-g)^{v} \! \left[ (g_{\mu \nu} q^{\mu \nu} )^2 - g_{\mu \nu} q^{\nu \rho} g_{\rho \sigma} q^{\sigma \mu} \right] , \label{interaction}
\ee
where $\kappa$ is the  interaction strength. To keep our discussion simple, we consider the case $u = 1/2$, but point out that our results remain unchanged for the choice $u=0$.  Then, the Einstein's equations derived by varying (\ref{bi-gravity-action-1}) and (\ref{interaction}) with respect to $g^{\mu \nu}$ and $q^{\mu \nu}$ are, respectively, given by
\bea
G_{\mu \nu}(g) + \frac{ \lambda_g}{L^2} g_{\mu \nu} = L^2 [ T^{g}_{\mu \nu} + T^{(mg)}_{\mu \nu} ],  \label{Einstein-eq-1} \\
G_{\mu \nu}(q) + \frac{ \lambda_q}{L^2} q_{\mu \nu} = L^2 [ T^q_{\mu \nu} + T^{(mq)}_{\mu \nu} ],  \label{Einstein-eq-2}
\eea
where $G_{\mu \nu}(g)$ and $G_{\mu \nu}(q)$ denote Einstein's tensors for each sector. Additionally, $T^g_{\mu \nu}$ and $T^q_{\mu \nu}$ are to be understood as the stress energy tensors sourced by $g$ and $q$, respectively, when both $\beta$ and $\kappa$ are nonvanishing. They are explicitly given by
\bea
T^g_{\mu \nu} &\equiv& - \frac{1}{L^4} \sqrt{\frac{-q}{-g}} \Big[ ( \beta - 2 \kappa \,  g_{\rho \sigma} q^{\rho \sigma}) g_{\mu \lambda} q^{\lambda \tau} g_{\tau \nu} \nn \\
&& \qquad \qquad \quad + 2 \kappa g_{\mu \lambda} q^{\lambda \tau} g_{\tau \sigma} q^{\sigma \rho} g_{\sigma \nu} \Big]  ,  \label{Einstein-1} \\
T^q_{\mu \nu} &\equiv&  \frac{1}{L^4} \Big[ (\beta - 2 \kappa g_{\rho \sigma} q^{\rho \sigma}  ) g_{\mu \nu} + 2 \kappa g_{\mu \sigma} q^{\sigma \rho} g_{\rho \nu} \qquad\qquad \nn  \\ 
&& \!\!\!\!\!\!\!\!\!\!\!\!\!\!\! - \frac{1}{2}  (\beta g_{\rho \sigma} q^{\rho \sigma} - \kappa (g_{\mu \nu} q^{\mu \nu} )^2 +  \kappa g_{\mu \nu} q^{\nu \rho} g_{\rho \sigma} q^{\sigma \mu}   ) q_{\mu \nu}  \Big], \label{Einstein-2}
\eea
whereas $T^{(mg)}_{\mu \nu}$ and $T^{(mq)}_{\mu \nu}$ represent stress energy tensors from matter fields coupled to each sector.

\section{Inflationary backgrounds}

We are interested in studying homogeneous and isotropic backgrounds. Disregarding intrinsic curvature effects, we may choose our two metrics to satisfy the following Ans\"atze consistent with these requirements
\bea
ds_{g}^2 = a^2 (- d \tau^2 + d {\bf x}^2), \label{homogeneous-metrics-1} \\
ds_q^2 = - X^2 d \tau^2 +  Y^2 d {\bf x}^2 , \label{homogeneous-metrics-2}
\eea
where $a = a(\tau)$, $Y=Y(\tau)$, and $X=X(\tau)$ are scale factors that depend only on $\tau$, the conformal time with respect to $g_{\mu \nu}$.  Notice that by a suitable change of coordinates, we can always find a frame where $q_{\mu \nu}$ is conformally flat instead of $g_{\mu \nu}$. For most of this discussion, we focus our attention on vacuum solutions whereby $T^{(mg)}_{\mu \nu} = T^{(mq)}_{\mu \nu} = 0$, and comment on the inclusion of matter fields later on.  We can already learn much from (\ref{homogeneous-metrics-1}) and (\ref{homogeneous-metrics-2}) by independently combining the $00$ and $11$ components of (\ref{Einstein-eq-1}) and (\ref{Einstein-eq-2}), respectively, leading to
\bea
\mathcal{H}'_a -  \mathcal{H}_a^2 = \frac{(X^2 - Y^2) (\beta Y^2 - 4 \kappa a^2 )}{2 L^2 X Y} , \label{background-eq-1} \\
\mathcal{H}'_Y -  \mathcal{H}_Y \mathcal{H}_X = - a^2 \frac{(X^2 - Y^2) (\beta Y^2- 4 \kappa a^2 )}{2 L^2 Y^4} ,  \label{background-eq-2}
\eea
where $\mathcal{H}_a= a'/a$, $\mathcal{H}_X= X'/X$, and $\mathcal{H}_Y= Y'/Y$ (here, primes $'$ denote derivatives with respect to $\tau$).
Now, the only way of achieving  vacua characterized by $\mathcal{H}'_a -  \mathcal{H}_a^2 = \mathcal{H}'_Y -  \mathcal{H}_Y \mathcal{H}_X= 0$ is either by having $X = Y$ or $Y = 2 \sqrt{\kappa / \beta} a$. Examples of such vacua are precisely Minkowski and de Sitter vacua. The first branch $X=Y$ corresponds to a case where both metrics are conformal to each other ($q_{\mu \nu} \propto g_{\mu \nu}$), whereas the second branch exists only if both $\kappa$ and $\beta$ are nonvanishing. To further understand these two branches we try the following scale factors representing two copies of de Sitter spacetimes
\be
a = - \frac{ L}{\tau H_0} , \quad X =  C_X a(\tau) ,  \quad Y =  C_Y a(\tau) , \label{scale-factors}
\ee
 (with $\tau <0$), where $H_0$ is a positive dimensionless constant determining the expansion rate $H \equiv H_0 / L$ of the homogeneous spacetime. Here,  $C_X$ and $C_Y$ are positive constants determining the second metric $q_{\mu \nu}$ in terms of $a(\tau)$. 
With (\ref{scale-factors}), both Eqs.~(\ref{background-eq-1}) and (\ref{background-eq-2}) are simultaneously reduced to $(C_X^2 - C_Y^2)( C_Y^2 \beta - 4 \kappa )= 0$. There are only two additional independent equations, given by
\bea
3 C_X  H_0^2  +  6 C_Y \kappa - C_X \lambda_g  - C_Y^3 \beta   = 0 , \label{C-eqs-2}\\
   (3 C_X^2 + C_Y^2 )  \beta   - 4 C_Y^2 (3 H_0^{2}  -  \lambda_q C_X^2) = 0 . \label{C-eqs-3}
\eea
We start by analyzing  the first branch $C_X = C_Y$, referred to as the proportional vacuum (PV). In this case, it is straightforward to solve (\ref{C-eqs-2}) and (\ref{C-eqs-3}) to obtain
\be
 H_0= \sqrt{\frac{\lambda_q \bar \lambda_g - \beta^2}{3(\lambda_q - \beta) }}, \quad
 C_X = C_Y =  \sqrt{\frac{\bar \lambda_g - \beta }{\lambda_q - \beta }} ,  \label{prop-sol}
\ee
where we have defined $\bar \lambda_g = \lambda_g - 6 \kappa$. Notice that Minkowski spacetimes may be obtained from this solution by tuning the parameters  to satisfy $\lambda_q \bar \lambda_g = \beta^2$. Additionally, in the limit $\kappa, \beta \to 0$ both metrics decouple and the solution reduces to standard de Sitter spacetimes for each gravitational sector. The second branch is called the nonproportional vacuum (NPV) and satisfies $C_Y =  2 \sqrt{ \kappa/\beta} $. It only exists for nonvanishing values of $\kappa$ and $\beta$; however, it may be reduced to a conventional de Sitter background by letting $\kappa, \beta \to 0$ with the ratio $\kappa / \beta$ fixed. We find it convenient to express $C_X$ and $H_0$ in terms of a single parameter $\theta$ as
\be
C_X = \theta C_Y ,  \quad C_Y  =  2 \sqrt{ \frac{\kappa}{\beta}}, \quad  3 H_0^2  = \lambda_g - 2 \frac{\kappa}{\theta}, \label{second-branch-eta}
\ee
where $\theta$ may be obtained by solving the following cubic polynomial equation:
\be
\left( 3 \beta + 16 \lambda_q \kappa /\beta  \right) \theta^3 + ( \beta - 4 \lambda_g) \theta +  8 \kappa  =0 .
\ee
Observe that in order to have a Minkowski vacuum $H_0 = 0$, one requires $\theta = 2 \kappa / \lambda_g$, and the previous equation implies that the parameters of the theory must satisfy $(12 \kappa^2 + \lambda_g^2 ) \beta^2 + 64 \kappa^3 \lambda_q = 0$. A remarkable property of the NPV is that the light cones of both metrics do not necessarily coincide. This means that each metric implies different maximal speeds to signals following their geodesics. While the maximal speed for particles following $g$-geodesics is normalized to be $c_g=1$, the maximal speed for particles following $q$-geodesics is $c_q=\theta$. (It was shown in~\cite{Blas:2005yk} that causality is preserved under these circumstances). Notice, however, that we could have worked in a frame where  $c_q = 1$ and $c_g = 1/\theta$.

\section{Scalar and vector perturbations}

We now study the evolution of perturbations on these backgrounds. We proceed by expanding the two metrics as $g_{\mu \nu} = g_{\mu \nu}^0 + h^S_{\mu \nu}+h^V_{\mu \nu}+h^T_{\mu \nu}$ and $q_{\mu \nu} = q_{\mu \nu}^0 +r^S_{\mu \nu}+r^V_{\mu \nu}+r^T_{\mu \nu}$ where $g_{\mu \nu}^0$ and $q_{\mu \nu}^0$ represent the background solutions of Eq.~(\ref{scale-factors}), and $S$, $V$, and $T$ denote scalar, vector, and tensor modes, respectively. Let us start by analyzing scalar perturbations. Since the theory is invariant under simultaneous gauge transformations on both sectors, we are then allowed to choose Newton's gauge to simplify one of the metrics, say $g_{\mu \nu}$, in the following way: 
\be
h^S_{00} =  - 2 a^2 \Psi , \quad h^S_{ij} = - 2 a^2 \Phi \delta_{ij} .
\ee
Since this gauge leaves no residual symmetry, $r^S_{\mu \nu}$ cannot be reduced in any similar way, and must be treated  in its most general form: $r^S_{0 0} = - 2 X^2 A$, $r^S_{0 i} = r^S_{i 0} = X Y \partial_i  F$,  and $r^S_{i j} =  Y^2  (-2 B \delta_{ij} + \partial_i \partial_j E )$, 
where $A$, $B$, $F$, and $E$ are the four scalar modes of the $q$-metric. By inserting these perturbations back into (\ref{Einstein-eq-1}) and  (\ref{Einstein-eq-2}) we deduce the linear equations for the evolution of all scalar modes. Many of the equations correspond to constraint equations which nontrivially couple the pair $(\Phi, \Psi)$ with $A$, $B$, $F$, and $E$.
In the particular case of the PV solution $C_X = C_Y$, the combination $\Sigma \equiv \Phi + \Psi$ decouples from the rest of the modes and satisfies the following equation of motion:
\be
\Sigma'' + \frac{2}{\tau} \Sigma' + k^2 \Sigma  +  M^2(\tau) \Sigma = 0, \label{Sigma-eq}
\ee
where $k$ labels the mode's wave number in Fourier space.
The mass $M^2(\tau)$ in Eq.~(\ref{Sigma-eq}) is found to be:
\be
\frac{M^2}{2 a^2} = \frac{ \beta + \lambda_q }{3 L^2  } C_X^2 + \frac{ 2 (\beta - \lambda_g + \lambda_q ) }{3 L^2  } +\frac{ 2 (\beta - \lambda_g)}{3 L^2 C_X^{2}  }.
\ee
Notice that the second term in Eq.~(\ref{Sigma-eq}) corresponds to a friction term with the opposite sign.
Because of this sign, after horizon crossing ($k < |\tau^{-1}|$) the combination $ \Phi + \Psi$ grows as $\Sigma \propto \tau^{-1} \cos (M \tau + \varphi)$ (where $\varphi$ is a phase determined by the initial conditions) regardless of the values of $\beta$, $\kappa$, $\lambda_{g}$ and $\lambda_q$, rendering PV backgrounds unviable to accommodate inflation. In the particular case where $\kappa = \beta = 0$, some of the constraint equations leading to (\ref{Sigma-eq}) disappear altogether, and one must perform the perturbation analysis again, obtaining the conventional result for de Sitter vacua in GR, whereby all the scalar fluctuations vanish.

The instability of the combination $ \Phi + \Psi$ was already found in \cite{Banados:2008fj} for the particular case $\kappa = 0$ and $\beta \neq 0$. There, it was speculated that with the inclusion of $S_{\rm int}$ of Eq.~(\ref{interaction}), it would be possible to cure this instability. As we have seen, this instability persists for the PV; however, the existence of NPV solutions for $\kappa \neq 0$ opens up the possibility of having new stable backgrounds for the propagation of perturbations. This is indeed what we find: in the case of the NPV background (\ref{second-branch-eta}), all scalar perturbations are constrained to vanish (satisfying constraint equations of the form $k^2 \Phi = 0$), meaning that this background is purely geometric. This is further corroborated by the fact that, after a similar analysis, one finds that vector perturbations $h^V_{\mu \nu}$ and $r^V_{\mu \nu}$ also vanish in this vacuum.

\section{Tensor perturbations}

The discussion of the previous section implies that the only viable vacuum to realize inflation is the NPV solution, where the only dynamical degrees of freedom are the two tensor modes $h^T_{ij}$ and $r^T_{ij}$, which we now study. To proceed we write $h^T_{ij} = a^2 \gamma_{ij}$ and $r^T_{ij} =  a^2 \sqrt{\theta} C_Y   \chi_{ij}$, with $\gamma_{ij}$ and $\chi_{ij}$ traceless and restricted to satisfy $\partial^i \gamma_{i j } =\partial^i \chi_{i j } = 0$. The factor $\sqrt{\theta}C_Y $ in front of $\chi_{ij}$ has been introduced to ensure that both fields $\gamma_{ij}$ and $\chi_{ij}$ have the same kinetic energy normalization. Disregarding indices, the equations of motion for $\gamma_{ij}$ and $\chi_{ij}$ are found to be
\bea
 \gamma ''  -  \frac{2}{\tau} \gamma'   + k^2 \gamma   -  \frac{C_{Y}^2}{  \theta }  \frac{  a^2 \beta  (\theta^2 - 1)}{ L^2}  \left[\frac{\sqrt{\theta}  \chi}{C_Y}- \gamma \right]  = 0 , && \quad\quad  \\
 \chi '' -  \frac{2}{\tau} \chi'  +  \theta^2 k^2   \chi  +    \frac{a^2  \beta  (\theta^2 - 1)}{  L^2} \left[ \chi - \frac{ C_Y \gamma}{\sqrt{\theta}}  \right] =  0 .  && \quad
\eea
This set of equations describes a coupled system of gravitons, of which, only one is massive. The nonzero eigenvalue of the mass matrix is given by 
\be
m_\sigma^2 = a^2 \beta  (\theta^2 - 1)( \theta + C_{Y}^2 )/ \theta L^2 , \label{m-sigma}
\ee
from which we obtain the restriction $\theta \geq 1$ if $\beta \geq 0$ and $\theta < 1$ otherwise. Observe that in the short wavelength limit $\tau k \gg 1$, $g$-gravitons propagate with speed $c_g = 1$, whereas $q$-gravitons propagate at $c_q = \theta$. However, for $k \lesssim m_{\sigma}$ both modes remain mixed, and we have to proceed carefully. The massless and massive modes, hereby denoted $\xi$ and $\sigma$, may be obtained from $\gamma$ and $\chi$ through a rotation as
\be
\xi = \frac{ \sqrt{ \theta} \, \gamma + C_{Y} \chi  }{\sqrt{\theta +  C_{Y}^2  }}  , \quad \sigma =\frac{ \sqrt{ \theta} \, \chi - C_{Y} \gamma }{\sqrt{\theta +  C_{Y}^2  }}  .
\ee
The equations of motion for $\xi$ and $\sigma$ are then given by
\bea
\xi''  - \frac{2}{\tau} \xi'  + k^2 A_{\xi \xi}  \xi + k^2 A_{\xi \sigma}  \sigma= 0 , && \quad \\
\sigma''  - \frac{2}{\tau} \sigma'  + k^2  A_{\sigma \sigma}  \sigma + k^2 A_{\sigma \xi}  \xi + m_{\sigma}^2 \sigma = 0 , &&
\eea
where $A_{\xi \xi} =  \theta (1 + C_Y^2 \theta) / ( \theta+C_Y^2 ) $, $A_{\sigma \sigma} =   (\theta^3 + C_Y^2 ) / (\theta+C_Y^2 )$, and $A_{\xi \sigma} = A_{\sigma \xi} =   \sqrt{\theta} (\theta^2-1)C_Y  / (\theta+C_Y^2 )$. The two fields continue to be coupled through their kinetic terms. However, if $m_{\sigma}^2(\tau) \gg 2 / \tau^2$, the massive graviton decays quickly before horizon exit, and the only relevant degree of freedom becomes the massless mode. One way of addressing this situation is by deducing an effective theory for the massless mode valid for the regime $k^2 \ll m_{\sigma}^2$. Following \cite{Achucarro:2010jv}, we find that  the massive field may be expressed in terms of $\xi$ as  $\sigma \simeq -   k^2 A_{\sigma \xi} \xi /(m_{\sigma}^2 - 2 / \tau^2 + k^2 A_{\sigma \sigma})$, and the dynamics of the massless mode is well described by
\bea
\xi''   - \frac{2}{\tau} \xi'  + k^2 \bigg[ A_{\xi \xi}  -  \frac{ k^2 A_{\sigma \xi}^2 }{m_{\sigma}^2 - 2 / \tau^2 + k^2 A_{\sigma \sigma}} \bigg] \xi  = 0 .  \label{masless-graviton} \qquad
\eea
This corresponds to a massless graviton with a modified dispersion relation. Notice that if the parameters of the theory $\lambda_{g}$, $\lambda_{q}$, $\beta$, and $\kappa$ are all of order unity, then the condition $m_{\sigma}^2 \gg 2/ \tau^2$ is equivalent to $H_0 \ll 1$ (or $H \ll L^{-1}$), which is necessary in order to trust the present field theoretical description of our system.  Since we are interested in phenomena for which $k^2 \ll m_{\sigma}^2$, the contribution coming from the second term inside the bracket in (\ref{masless-graviton}) may be neglected, and the speed of propagation $c_h$  for this mode becomes
\be
c_h^2 =  \theta (1 + C_Y^2 \theta) / ( \theta+C_Y^2 )  . \label{graviton-speed}
\ee
Notice that $c_{h} \in [1,\theta]$, depending on the value of $C_Y^2 = 4 \kappa / \beta$. This is because the massless mode appears from the mixing between both metrics, and therefore its propagation is affected by both backgrounds simultaneously.  

One crucial aspect of the results summarized in Eqs.~(\ref{masless-graviton}) and~(\ref{graviton-speed}) is that even for a length scale $L$ of the order of the Planck length scale, the difference between the two light cones can be large, and therefore the mixing between the two tensor modes $\gamma$ and $\chi$ may be sizable. This would translate in a speed of sound $c_h$ considerably different from the two values $c_g = 1$ and $c_q = \theta$. 

To finish this section, we notice that the cutoff scale determining the validity of the effective description for the massless graviton $\xi$ in Eq.~(\ref{masless-graviton}) is given by the mass $m_\sigma$ of the massive graviton, given in Eq.~(\ref{m-sigma}). That is, for energies below the cutoff energy scale
\be
\Lambda \sim m_\sigma ,
\ee 
we recover a theory of fluctuations propagating in a de Sitter background where only one graviton is in charge of propagating the gravitational force. This in turn means that for energies below $\Lambda$, we may study the evolution of perturbations with the help of standard effective field theory techniques, consistent with the symmetries of the background. We shall exploit this fact in the next section, where we study the phenomenological consequences of our results.

\section{Consequences}

The previous results have some interesting and nontrivial consequences that we now discuss. To achieve realistic models of inflation, we need to move our analysis from de Sitter to quasi-de Sitter spacetimes, in such a way that the background quantities $H_0$, $C_X$, and $C_Y$ evolve adiabatically towards Minkowski. This may be achieved, for instance, by introducing a scalar field $\phi$ into the theory, and letting  $\lambda_{g}$, $\lambda_{q}$, $\beta$, and $\kappa$ be functions of it. Then, the slow roll of $\phi$ towards a Minkowski vacuum (where $H_0=0$) would make the background depart slightly from the de Sitter configuration analyzed in the previous sections. The dependence of these parameters on $\phi$ will be restricted to satisfy certain slow-roll conditions, just as in the case of conventional slow-roll inflation. Another more interesting and challenging way of obtaining inflation would be without the assistance of a scalar field $\phi$, in which case the system offers a time dependent solution close to the de Sitter backgrounds previously discussed, where all the parameters $\lambda_{g}$, $\lambda_{q}$, $\beta$, and $\kappa$ stay constant. In the present discussion, we disregard these model dependent aspects related to the background, and focus only on the perturbations of the theory.

To start with, since in pure de Sitter there are no scalar degrees of freedom, in this new quasi-de Sitter phase there will necessarily exist a comoving curvature mode $\zeta$ or, equivalently, a Goldstone boson mode~\cite{Cheung:2007st}, reflecting the fact that time translation symmetry has been broken. Because we assume this time translation to be slightly broken by the background dynamics, we expect a nearly scale invariant power spectrum of scalar perturbations. The specific form of such a power spectrum will depend on a number of details out of the scope of the present article, such as the number of additional scalar modes interacting with the curvature mode $\zeta$, and we leave this question open for future work. 

The novel aspect comes with tensor modes:  since they propagate with a variable speed $c_h$ in the regime $c_h^2 k^2 \ll m_{\sigma}^2$, the power spectrum of tensor modes departs from the conventional prediction encountered in GR. To arrive at concrete predictions, we assume that matter fields are coupled to $g_{\mu \nu}$ and $q_{\mu \nu}$ in such a way that it only couples to the massless mode $\xi$~\cite{Clifton:2010hz}. Other alternatives may be considered but would lead to similar conclusions. Then, by imposing Bunch-Davis vacua for subhorizon modes (in the regime $a^2 H^2 \ll k^2 \ll m_{\sigma}^2$) we find:
\be
\mathcal{P}_{T} (k) = \frac{2 L^2 H^2}{\pi^2 c_h^3} (k/k_0)^{n_T} . \label{tensor-spectrum}
\ee
In the particular case where the scalar sector could be effectively described by a conventional scalar field theory, Eq.~(\ref{tensor-spectrum}) would modify the tensor to scalar ratio to be
$
r = 16 \, \epsilon \, / c_{h}^3 ,
$
where $\epsilon = - \dot H / H^2$ (here, $\dot{}$ denotes a derivative with respect to cosmic time $dt = a d\tau$).
Additionally, the spectral index $n_T$ of tensor perturbations is now found to be
$
n_T =  - 2 \epsilon - 3 \epsilon_h ,
$
where $\epsilon_h \equiv  \dot c_h  / H c_h$. Since $c_h$ is sensitive to the specific dependence of $\lambda_{g}$, $\lambda_{q}$, $\beta$, and $\kappa$ on $\phi$, the power spectrum $\mathcal{P}_T(k)$ in bigravity models is no longer restricted to be red tilted. Finally, the effects of the UV field $\sigma$ on $\mathcal{P}_T(k)$ are of order $H^2/m_{\sigma}^2 \ll 1$. Notice that the specific result (\ref{tensor-spectrum}) depends on the specific choice for the coupling between the tensor perturbations and matter fields (where matter fields only couple to the massless mode $\xi$). Other more general couplings will only modify this prediction by changing the effective value of the Newtonian constant in terms of the fundamental length scale $L$ (here given by $G_N = L^2 / 8 \pi$) and therefore will only change the amplitude of (\ref{tensor-spectrum}).

\section{Conclusions}

We see that the realization of inflation severely constrains the parameter space of bigravity models. Our main result is summarized in Eq. (\ref{masless-graviton}), which shows that tensor modes propagate with a modified dispersion relation. As a consequence, the predicted power spectrum for tensor modes differs from the conventional prediction offered by slow-roll inflation. In addition, we expect other relevant departures from the conventional picture such as the enhancement of non-Gaussian distribution of tensor modes (see also~\cite{Maldacena:2011nz}). The importance of these results is twofold. On the one hand, it forces us to widen our view of effective theories of inflation to include consequential modifications to the tensor sector of the theory, consistent with the symmetries of quasi-de Sitter spacetimes~\cite{Cheung:2007st, Weinberg:2008hq}. On the other hand, it reemphasizes the need for improving the precision of CMB polarization measurements~\cite{Baumann:2008aq} in order to further test the physics of the very early Universe. Since primordial B-mode signals are exclusively due to tensor perturbations, the next generation of CMB polarization probes might give a powerful insight into bigravity theories. Last but not least, notice that Eq.~(\ref{masless-graviton}) is also valid for Minkowski spacetimes, and therefore it implies significant levels of departures from GR for gravitational wave phenomenology. For instance, $c_h > 1$ would involve a travel delay of gravitational waves from their sources when compared to light signals. Otherwise, $c_h < 1$ would produce a Cherenkov type of radiation emitted by particles exceeding $c_h$.

\begin{acknowledgments}
We would like to thank Max Ba\~nados, Richard Easther, Walter Gear, and Gustavo Niz for useful comments and discussions.  This work was partially supported by the Center of Excellence in Astrophysics and Associated Technologies (PFB 06) (VA \& LEC), by the Centro de Astrof\'isica FONDAP 15010003 (VA), by a Fondecyt Iniciaci\'on Project No. 11090279 (GAP), and by a Conicyt Anillo Project No. ACT1122.
\end{acknowledgments}

\end{document}